# *CostMAP*: An open-source software package for developing cost surfaces

**Brendan Hoover**
Earth and Environmental Sciences
Los Alamos National Laboratory

**Richard S. Middleton**
Earth and Environmental Sciences
Los Alamos National Laboratory

**Sean Yaw**
Gianforte School of Computing
Montana State University

March 27, 2019

## Abstract

Cost Surfaces are a quantitative means of assigning social, environmental, and engineering costs that impact movement across landscapes. Cost surfaces are a crucial aspect of route optimization and least cost path (LCP) calculations and are used in a wide range of disciplines including computer science, landscape ecology, and energy infrastructure modeling. Linear features present a key weakness to traditional routing calculations along costs surfaces because they cannot identify whether moving from a cell to its adjacent neighbors constitutes crossing a linear barrier (increased cost) or following a corridor (reduced cost). Following and avoiding linear features can drastically change predicted routes. In this paper, we introduce an approach to address this adjacency issue using a search kernel that identifies these critical barriers and corridors. We have built this approach into a new Java-based open-source software package—CostMAP (cost surface multi-layer aggregation program)—which calculates cost surfaces and cost networks using the search kernel. CostMAP not only includes the new "adjacency" capability, it is also a versatile multi-platform package that allows users to input multiple GIS data layers and to set weights and rules for developing a weighted-cost network. We compare CostMAP performance with traditional cost surface approaches and show significant performance gains—both following corridors and avoiding barriers—using examples in a movement ecology framework and pipeline routing for carbon capture, and storage (CCS). We also demonstrate that the new software can straightforwardly calculate cost surfaces on a national scale.

# 1 Introduction

In this paper, we present a new multi-scale method to determine whether adjacent raster cells that contain line features act as barriers or corridors in costs surfaces. We demonstrate our new method with two representative case studies—wildlife-movement ecology and CCS infrastructure. First we illustrate a movement ecology framework because biologically realistic cost surfaces are crucial for modeling and understanding animal movements (Sawyer, Epps and Brashares 2011). Second, we demonstrate that realistic cost networks can vastly improve infrastructure decisions for commercial-scale CCS, both the underlying pipeline routing, as well as the decision of where to capture and store $CO_2$. As part of this research, we also introduce an open-source software called the Cost Surface Multi-layer Aggregation Program (*CostMAP*), which utilizes our multi-scale method for straightforwardly and flexibly weighing multiple geographic features as well as incorporating linear features as barriers or corridors.

*CostMAP* is a flexible, multi-platform, open-source software package which calculates weighted-cost surfaces and weighted-cost networks. The cost surface is a raster image that can be used in any geographic information system (GIS) software. The cost network is a file that includes the costs of moving between every cell, which can be utilized in scripting languages like python or R, and is used by *SimCCS$^{2.0}$*, an economic-engineering framework designed to find optimal network solutions for carbon capture, and storage (CCS) (Middleton et al. 2018, Middleton and Bielicki 2009). The Java-based *CostMAP* runs in Mac, Linux, and Windows operating systems, and is available as a standalone product or as part of *SimCCS$^{2.0}$*. The cost surfaces and cost networks generated by *CostMAP* can be used in many ways—such as siting facilities—but were originally designed for Least Cost Path (LCP) analysis.

Many different geographic features or attributes—such as terrain, land cover, land ownership, and population density—act as essential variables in determining LCPs (Miller 2004). Typically, GIS programs, like GRASS, QGIS, or ArcMap/ArcGIS Pro, calculate costs as an additive sum of geographic and social factors. *CostMAP* accounts for similar factors, but uses a combination of accumulation, binary selection, and proportional weighting. For example, land cover weights are overridden at certain thresholds of population density. *CostMAP* also includes a new and transformative approach to incorporating barriers (such as rivers and roads) and corridors (such as existing rights of way). The ability to weigh linear features as either barriers or corridors significantly increases cost accuracy and more realistically predicts routes than is typically possible using traditional approaches. The improved representation of corridors in *CostMAP* essentially allows users to represent existing 2D vector networks (existing corridors such as pipelines) as well as identify new low-cost routes (a new pipeline would use an existing ROW *if cheaper*) within a single surface, addressing needs such as entering and leaving the 2D network at any point (Choi at al.)

A key unaddressed challenge to applying weights to cost surfaces arises when identifying whether cells that contain line features—like rivers, roads, or pipelines—constitute a barrier, corridor, or both. This challenge is an issue of scale that occurs when movement between cells is limited to adjacent cells with the same resolution. When a cell contains both a barrier and a corridor, the 'from cell' and 'to cell' connection actually defines whether a route is crossing a barrier or leveraging a corridor. For example, in Figure 1 it is possible to move adjacently from the centroid of cell B to the centroid of cell C without crossing the river, but traditional cost surface calculations in GIS would weigh movement between the cells as if the river were being crossed twice (Fera 2007). Similarly, the adjacent movement between cells B and E might be aided by a river corridor, whereas adjacent movement between cells A and B would not.

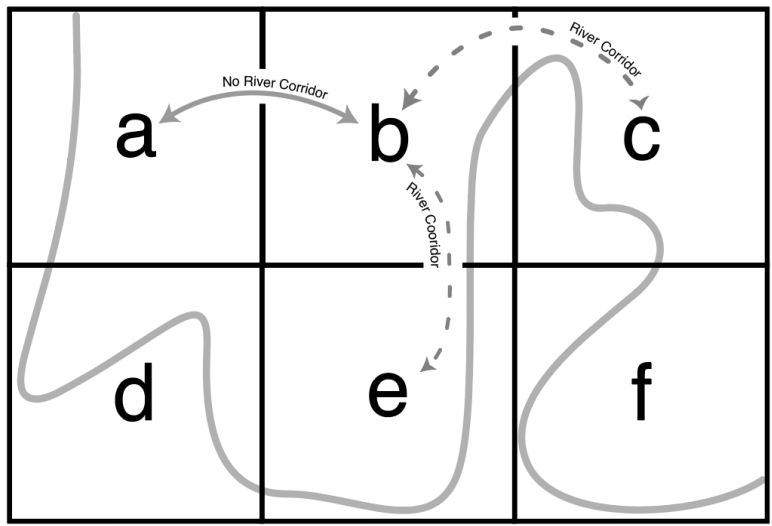

*Figure 1: Illustration of the adjacency issue when two cells contain a river. The river enters all six cells, but some of the cells are still connected, such as cells B and E, without having to cross the river.*

## 2 Background

## 2.1 Least cost paths

LCP analysis provides essential insights into many social and scientific issues. For instance, LCP analysis finds routes for transportation, pipelines, transmission lines, and information technology. LCP analysis is also beneficial to climate change mitigation in terms of developing large-scale CCS pipelines and other infrastructure (Middleton et al. 2018, Middleton and Bielicki 2009); this one of our two case studies. Ecologists use LCP to understand gene flow, biodiversity (Urban and Keitt 2001), and how animals utilize their landscape (Sawyer et al. 2011); this is the basis for our second case study. To understand historical trade routes and city structures, anthropologists regularly utilize LCP analysis (Bicho, Cascalheira and Goncalves 2017). LCP analysis has been used to find solutions to reduce city congestion (Wen, Çatay and Eglese 2014). Even the tourist trade can utilize LCP to determine optimal tours for scenic or hidden routes (Stucky 1998).

Euclidian distance defines the shortest path between point pairs in homogeneous spaces, but the shortest path in heterogeneous spaces, like transportation networks or environmental landscapes, is defined by distance as well as geographic and social factors. Graph theory, which is a branch of mathematics that analyses the pairwise relationship of point objects, has been successfully used to find shortest paths that account for factors beyond Euclidian distance (Evans 2017). For instance, graph theory has been widely used in a variety of disciplines which are concerned with maximizing routing and flow efficiency, such as computer science, urban planning, and landscape ecology (Urban and Keitt 2001).

Graph theory algorithms calculate the shortest path between vertices or nodes by minimizing the weights of connecting lines, which are referred to as edges. Hundreds of algorithms have been developed to solve shortest path problems (Deo and Pang 1984), but the most widely used in GIS applications is Dijkstra's shortest path algorithm (Dijkstra 1959). Dijkstra's algorithm solves the shortest-path problem on an edge-weighted graph, where all edge weights are non-negative. Dijkstra's algorithm utilizes a breadth-first search to explore all adjacent nodes in a graph before moving onto non-adjacent nodes (Evans 2017). Given a graph $G = (V, E)$ and a source vertex $s$, a breadth-first search explores all edges in $G$ to discover all the vertices that are reachable from $s$. The shortest path corresponds to the lowest weight of accumulated edges (Cormen et al. 2009).

In GIScience, edge weights in graphs are typically calculated via raster-based accumulated cost surfaces, which quantify the cost of moving across grid cells by combining social and environmental factors. Along with these accumulated environmental and social factors, optimal least cost paths also account for the minimized distance between node pairs. In the case of rasters, distance and angle is typically consistent across a graph, which makes raster-based graphs a special case (Hopkins 1973). The center of each cell in the raster-cost surface is defined as a node, and the line segments between two nodes as edges (e.g., Huber and Church 1985). A stepwise process computes edge weights as the cost of moving from cell to cell via rook's or bishop's kernels (named from the movement of chess pieces). Both rook's and bishop's movements are computationally non-intensive but create unnatural stair step paths known as proximity distortions because movement between cells is limited to adjacent cells (Huber and Church 1985).

The queen's kernel (Figure 2) which is a combination of rook's and bishop's movement, is a compromise to computational speed, proximity distortion, and is the most common used in GIScience, but is still impacted by the limitations of cell adjacency. When using a queen's kernel, if an adjacent cell contains a linear feature like a road, the standard procedure is to increase that cell's cost whether or not the barrier has been crossed. Real world illustrations demonstrate the importance of weighing cell barriers more accurately. For example, mortality risks increase for wolves when they cross roads because of automobiles (Merrill and Mech 2000), but roads can also aid wolf movement when they hunt along or near their edges (Whittington et al. 2011). Similarly, when building pipeline infrastructure, roads can act as both barriers and corridors (referred to as easements or rights of way (ROWs) in transport infrastructure terminology that includes legal designations). It is significantly more expensive to build a pipeline beneath a road to traverse the barrier but building a pipeline along a road is less expensive since many of the costs of creating the ROW (e.g., obtaining land rights, leveling topography) were already incurred during road construction. The easement for ROWs also enhance access to the pipeline during and after construction (Lugschitz 2017). Increased accuracy in the calculation of barrier and corridor weights will increase the accuracy of accumulated cost surfaces and the conclusions they support.

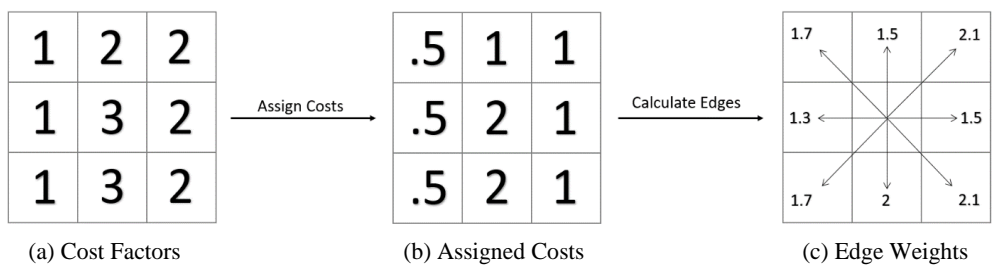

(a) Cost Factors      (b) Assigned Costs      (c) Edge Weights

*Figure* 2. *Illustration of assigning edge costs via a queen's kernel: (a) shows the inputted values for the accumulated costs of construction, environmental, and social factors; (b) these factors are then assigned costs; (c) the edge weights are calculated based on distance and assigned costs.*

## 2.2 *CostMAP*

To implement the search kernel, we developed a software program called the Cost surface multi-layer aggregation program (*CostMAP*) which builds weighted-cost surfaces and weighted-cost networks via inputs from a user interface. *CostMAP* includes pre-processed default data and weights for land cover, slope, aspect, population density, protected lands, rivers, roads, railways, and pipeline networks, but allows users to use their own data and to weigh these factors differently. Users are able to use fewer than the default number of layers, as well as specify entirely additional layers. This is particularly useful if the user has data that are not publicly available such as land values or transmission lines. Further, the open-source nature of *CostMAP* means that users can even edit the code to account for new situations. Ultimately, *CostMAP* is designed to increase the accuracy of barriers and corridors, while enhancing scientific study through a flexible software approach.

To accurately identify barriers and corridors, the search kernel first checks if a cell contains a linear feature (such as a river or road), then uses a more fine-scale raster to carefully check for actual crossings or corridors. Existing approaches are typically limited to a cell being a crossing *or* a barrier if any part of the cell is a barrier or corridor (regardless if there is actually a barrier or corridor) and cannot ever have a cell be both a barrier and corridor as exists in reality. We refer to the cells in the coarser-scale raster as "Major Cells" and those in the finer-scale raster as "Minor Cells". For barriers, the challenge lays in identifying which Major Cells are connected without having to cross the line features that are present in the Minor Cells. Major cells contain a crossing based upon the configuration of linear features within the Minor Cells. Figure 3 illustrates six major raster cells and a single line feature (a river) that runs through every Major Cell. In a traditional cost surface approach, all six cells would be identified as a crossing because the river appears in every cell. Further, moving between any of the 10 possible internal adjacencies (as well as all external links) would incur a crossing penalty. However, Figure 3a shows that of the ten internal adjacencies (i.e., between the Major Cells A through F), five adjacent cells can be connected without having to cross the river. These are rules created and implemented in *CostMAP* that contain all possible cases where crossings could occur between the Minor Cells within two adjacent Major Cells (Figure 3a). As a result, unlike traditional approaches, our new approach is not over-estimating costs incurred by barriers. And, in cases where barriers are entirely prohibitive, *CostMAP* is able to find feasible routes that other approaches cannot identify.

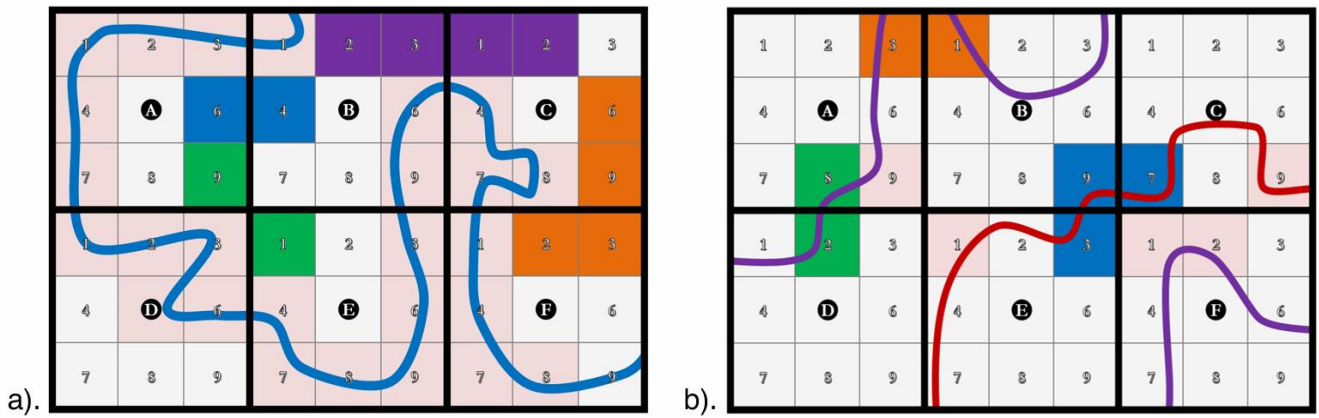

*Figure 3(a). Illustration of identifying crossings using the search kernel. Thick black lines are Major Cells (lower-resolution). Thin grey lines are Minor Cells (higher-resolution). The blue line is a river feature that forms a crossing barrier. Minor Cells that are "closed" (i.e., Minor Cells that contain the river) are shaded light red. Minor Cells that are not shaded are "open" for unencumbered movement. The vibrantly-colored Minor Cells (blue, green, purple, and orange) illustrate the different ways two Major Cells can be connected without having to cross the river. 3(b). Illustration of checking for corridors using the search kernel. The purple (roads) and red (pipeline) lines are linear features that potentially increase connectivity between Major Cells. The colored (blue, green, and orange) Minor Cells represent potential corridors that connect two Major Cells.*

The same principles were applied to create a set of rules to identify which Major Cells can be connected as potential movement corridors (Figure 3b). Existing ROWs can dramatically reduce costs and increase connectivity, and thus are critical to identify when using cost surfaces for lowest-cost routing. Similar to the barriers example, all six Major Cells in Figure 3b contain either a pipeline and/or a road ROW. In a traditional cost surface approach, all ten possible internal adjacencies would be counted as a corridor connection and therefore have their weights unrealistically reduced. In reality, only five such adjacencies exist in this example (Major Cell pairs AD, AB, BE, BC, and CE in Figure 3b). For two Major Cells to be connected by a corridor, the line feature has to be present in two adjacent Minor Cells. This can include cases where the line feature does not precisely connect two adjacent Minor Cells (e.g., A3 and B1 in Figure 3b ensure Major Cells A and B are connected) but does not include adjacent Minor Cells of different ROW types (e.g., Minor Cells E3 and F1 do not constitute a connection between Major Cells E and F because one is a pipeline and the other road). This improved ability to identify corridors and ROWs is critical for accurately identifying appropriate routes and calculating costs. Although not shown, *CostMAP* seamlessly calculates barriers and corridors for all situations, including where cells have both corridors and barriers.

In addition to more realistically capturing the impact of barriers and corridors, *CostMAP* develops cost surfaces that capture the non-linear interaction between multiple geographic variables as opposed to simple cumulative calculations. For example, *CostMAP* includes dynamic weighting of populated areas so that less dense areas are favored over more dense regions. In this case, the population weights can be used to override weights calculated from variables such as land ownership or land cover (i.e., binary choices, not accumulated costs). *CostMAP* also includes more complex operations such as adjusting costs according to change in slope and aspect between cells; this can be used to avoid or increase costs for constructing pipelines up and down slopes, but takes into account aspect so that construction can run parallel to the slope unimpeded. The cost surface tool also includes weighting to avoid protected lands such as national parks.

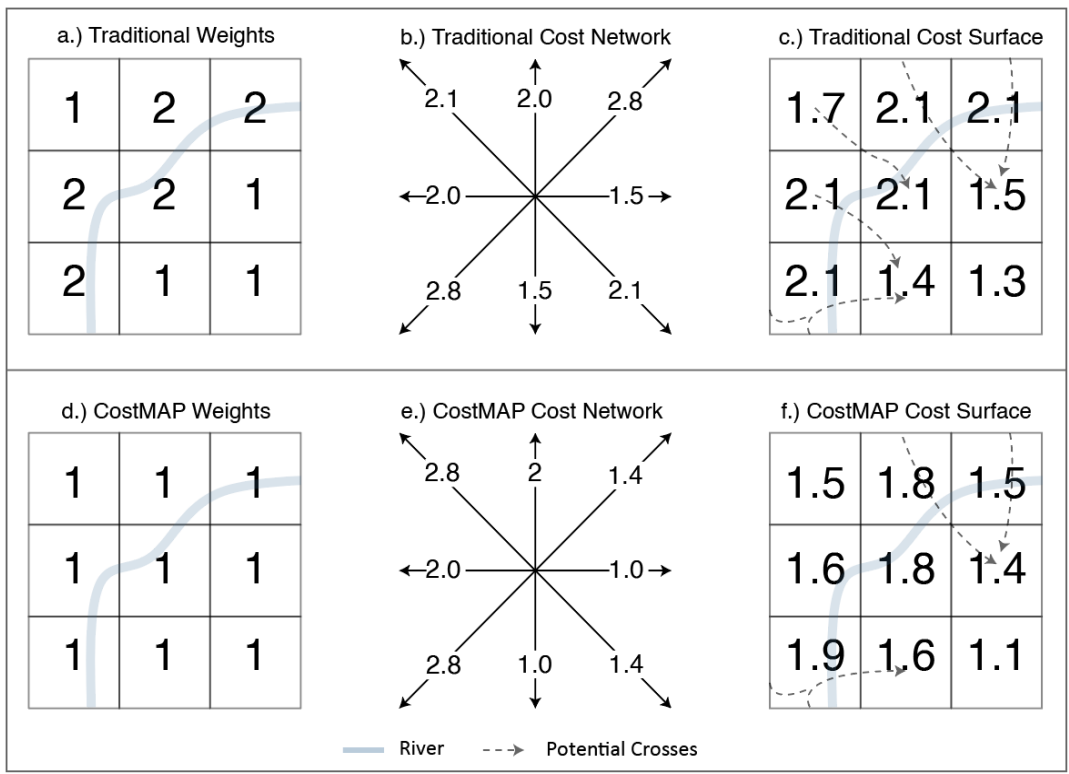

*Figure 4. An example of the aggregated cell weights due to a river crossing using a traditional queen's kernel and the search kernel. (a/d) Shows the weights of each cell before calculation using a traditional versus CostMAP weights where the land cover is uniform and weights are adjusted due to the river. (b/e) Shows the cost network calculations from those weights and the crossing of the river. (c/f) Shows the cost surfaces and includes potential river crossings assuming entering a cell from all potential directions and movement to a lower valued cell.*

*CostMAP* outputs a weighted-cost network with the cost of moving between every adjacent cell pair in the domain. In addition, this cost network is collapsed to produce an aggregated cost raster, where up to all eight connections to neighboring cells are collapsed into a single cost for each raster cell (cell values for the aggregated cost raster are calculated as the mean of all adjacent nearest neighbors to the centroid cell). This is done for visualization purposes and for applications that cannot take advantage of the networking. Figure 4 shows an example of the difference in aggregation values for a cell using the traditional queen's kernel (Figure 4c) and the search kernel in *CostMAP* (Figure 4f), as well as a visualization of the cost networks (Figure 4b/e).

Using the traditional cost surface, the cell values are doubled if a river is present (Figure 4a). Consequently, in the traditional cost surface network (Figure 4b), the top left cell would not be weighted for a river crossing when it should not be. Using our new search kernel method (Figure 4e), the top left cell is weighted appropriately for a river crossing. Though the values are aggregated in Figures 4c and 4f, the relative cell weights are more reflective of the river crossing, which prevents unnecessary crossings more often. Since the aggregated cost surface is a raster file, it is straightforward to use in any GIS-aware software. The cost network (Figure 4e) shows the value of moving to each cell from the center cell. The cost-network output can be used for LCP calculations in any software; the output file is by default set to be read into *SimCCS*$^{2.0}$, but could be easily modified for any other software or via scripting languages such as R or Python.

# 3 Movement ecology case study

## 3.1 Data and methods for movement ecology case study

Cost surfaces are often an instrumental part of animal movement analysis and are commonly used to understand how terrestrial animals move in relation to their environment (Rayfield, Fortin and Fall 2010). For example, cost surfaces have been used in the analysis of Eurasian Lynx (*Lynx lynx*) (Schadt et al. 2002), Hedgehogs (*Erinaceuse erupaeus*) (Driezen et al. 2007), Easter grey squirrels (*Sciurus caolinensis*) (Gonzales and Gergel 2007), and Florida Panthers (*Puma concolor corryi*) (Kautz et al. 2006). To evaluate *CostMAP* cost surfaces in analyzing terrestrial animal movement, we created two scenarios related to movements near rivers. Rivers often act as barriers or corridors for terrestrial wildlife. In both scenarios, we used biased correlated random walks (BCRWs) to simulate the movements of Baird's tapirs (*Tapirus bairdii*) near rivers. We chose to simulate Baird's tapirs because they often move long rivers (Jordan et al. 2019) and have a large perceptual range, allowing them to perceive landscape structure (Garcia et al. 2012). We used simulated data because it allowed us to control the persistence levels of the data. This was important for distinguishing movement phases across the surfaces and allowed us to control and test groups for the statistical analysis. We used neutral landscapes modesl (NLM) to create a neutral landscape so we could focus on the tapir response to the river. In the first scenario, we made river crossings very costly to tapir movement. We compared how often the tapir crossed rivers over a traditional cost surface versus a cost surface created in *CostMAP*. In the second scenario, we made river cells very beneficial to tapir movement and compared how often the tapir used the river as a corridor in a traditional cost surface versus in a cost surface created in *CostMAP*. For both scenarios, we also compared the results of the tapir movements over traditional and *CostMAP* cost surfaces to a Null surface, which weighed movements as neutral to rivers.

### 3.1.1 Simulated landscape and river system

To assess the two river scenarios, we created a NLM using the NLMpy tool in Python (Etherington, Holland and O'Sullivan 2015). NLMs are a classic tool for examining ecological patterns and processes across landscapes. Their use in movement ecology spans a wide array of questions, including those related to group dynamics (Langrock et al. 2014), dispersal (Lowe and McPeek 2014), and epidemiology (White, Forester and Craft 2018). NLMs are a valuable tool for understanding movement patterns because researchers can control the level of spatial dependence (With, Gardner and Turner 1997). In our case, spatial dependence is related to a river network. We generated an NLM raster with a 15 km$^2$ extent (consistent with tapir home-range size) with a resolution of 30 meters (Jordan et al. 2019). We used a random elements NLM which uses nearest-neighbor interpolation to create irregularly shaped but relatively consistent size patches that allows for the integration of line networks (Gaucherel 2008). The effect of using the NLM in this manner allowed us to create a river network that is similar to the habitat found by Baird's tapir. Using ArcMap 10.5, we generated line features for the river network using the NLM as a digital elevation model and calculating flow direction and flow accumulation.

### 3.1.2 Derived cost surfaces for comparison

Using the NLM, we derived five different cost surfaces for the barrier and corridor scenarios. In both scenarios, the first cost surface, which we refer to as the Null Surface, weighted the rivers as neutral to movement costs. For the barrier scenario, crossing rivers was weighted as an impediment to movement and two cost surfaces were created. First, we created a cost surface using a traditional queen's kernel, which we refer to as Traditional-Surface A. Then we created a cost surface using the search kernel in *CostMAP*, which we refer to as *CostMAP*-Surface A. For the corridor scenario, we weighted rivers corridor as beneficial to movement and created two costs surfaces, first using the traditional queen's kernel (Traditional-Surface B), and then using the search kernel available in *CostMAP* (*CostMAP*-Surface B).

### 3.1.3 Biased-correlated random walks

To evaluate movement across our simulated cost surfaces, we mimicked the movement of Baird's tapir using BCRWs. Each walk was biased toward low costs cells using the SimRiv package in R (Quaglietta and Porto 2018). The simulated tapir movements were recorded every 30-meters 10,000, which is consistent with tapir movements found in real-world telemetry data (Jordan et al. 2019). Tapir' step direction was also correlated with the previous step at different levels of persistence (Codling, Plank and Benhamou 2008). The BCRWs used concentration parameters of 0 (no directional persistence), 0.5 (medium directional persistence), and 0.9 (strong directional persistence). These different parameters allowed us to also consider how short-term goals (no directional persistence) and long-term goals (strong directional persistence) of the tapir could be impacted by the different cost surfaces. The movement of 1,000 tapirs at each persistence level were simulated for both the barrier and corridor scenarios.

### 3.1.4 Null model

To compare our search kernel with that of a traditional queen's kernel, we used a null model approach. This involves generating a random pattern which lacks the process or mechanism being tested (Miller 2015). For scenario 1, the process being tested was how often tapir would cross a river when crossings were weighed as negative versus when crossings were not. Our null hypothesis was that tapir river crossings would not be significantly different using the three different cost surface models (Null, Traditional-Surface A, and *CostMAP*-Surface A). To test this hypothesis, using each cost surface model, we counted the number of times a river was crossed for each simulated tapir.

For scenario 2, where the river acted as a corridor to movement, we tested how often a tapir would use the rivers as a corridor. Our null hypothesis was that the number of times a tapir used a river corridor would not be significantly different using the three different cost surface models (Null Surface, Traditional-Surface B, and *CostMAP*-Surface B). To evaluate this second scenario, we created a buffer around the river at 60 meters and counted how often each tapir used the river as a corridor. We considered the tapir using a river corridor when three consecutive steps were within the river buffer.

## 3.2 Results for movement ecology case study

For both river scenarios, our results did not meet the assumptions of parametric analysis of variance based upon QQ plots and the analysis of variance test for normality (Shapiro and Wilk 1965). Consequently, we used the Kruskal-Wallis statistic to test for significant differences between the results on the three costs surface models (Taylor, Nudds and Thomas 2003). Using the Wilcoxon-Mann-Whitney statistic (Cunha and Vieira 2002), we were able to test for significant differences between each of the three cost surface models for both the river crossing and river corridor scenarios.

### 3.2.1 Results for tapir avoiding river crossings

*Table 1. Wilcoxon-Mann-Whitney Comparison of p-values and z scores of 1,000 simulations of tapir movements at three different persistence levels, moving over three raster models (Null Surface, Traditional-Surface A, and CostMAP-Surface A). We used a probability level < 0.05 where the minimum score is equal to 0.001.*

| **Surfaces:** | **Null surface:** Movement not influenced by the river. | **Tradition A surface:** Movement weighted to avoid rivers using the queen's kernel. | *CostMAP* **A surface:** Movement weighted to avoid rivers using the search kernel |
|---|---|---|---|
| **Comparison** | Tapir (0.0) | Tapir (0.5) | Tapir (0.9) |
| **Traditional A – *CostMAP* A** | 0.001 (z = -11.09) | 0.001 (z = -15.07) | 0.001 (z = -9.06) |
| **Traditional A – Null Surface** | 0.8238 (z = 0.93) | 0.001 (z = -25.77) | 0.001 (z = -19.20) |
| ***CostMAP* A – Null Surface** | 0.001 (z = -10.46) | 0.001 (z = -27.26) | 0.001 (z = -19.34) |

With a p-value of 0.001 the Kruskal-Wallis test showed a significant difference at a < 0.05 probability level for all tests in the river crossing scenario (Table 1). The z-scores are also reported to indicate the magnitude of the difference between the tests for each raster model.

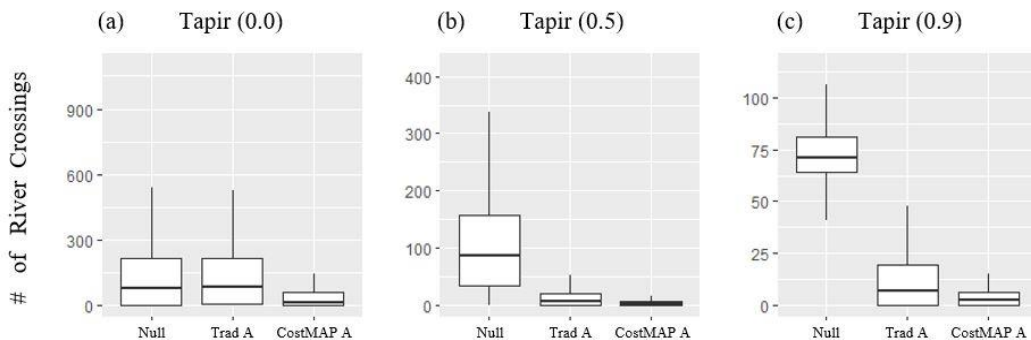

*Figure 5. Boxplots of the number of times the tapir crossed a river when walking over each raster model (Null Surface, Traditional-Surface A, and CostMAP-Surface A). The plots show the five-number summary (min, max, standard deviation, range and mean) of the number of times a river was crossed by 1,000 individual tapirs at different persistence levels.*

The boxplots in Figure 5 visualize the differences between the three raster surfaces and how tapirs responded to the different weightings of the cost models. As the persistence level increased from 0.0 to 0.9, there were fewer river crossings in all scenarios. The *CostMAP* surface prevented river crossings the most for all simulations. In the case of the tapir with a 0.5 and 0.9 persistence level, the surface created with the queen's kernel also performed better at preventing river crossings than the Null surface. However, for tapir with a 0.0 persistence level, the number of crossings on that surface were not significantly different than on the Null Surface. The relatively high z-scores suggest there is a strong difference between how often tapir crossed rivers on each surface. This result also indicates that *CostMAP* surfaces are much better at preventing crossings than traditional surfaces.

### 3.2.2 Results for tapir using a river to aid movement

*Table 2. Wilcoxon-Mann-Whitney Comparison of p-values and z scores of 1,000 simulations of tapir walks for the river corridor scenario and for the three-raster model (Null Surface, Traditional-Surface B, and CostMAP-Surface B). We used a probability level < 0.05 where the minimum score is equal to 0.001.*

| **Surfaces:** | **Null surface:** Movement not influenced by the river. | **Tradition B surface:** Movement weighted to use river as cooridor using the queen's kernel. | ***CostMAP* B surface:** Movement weighted to use river as cooridor rivers using the search kernel |
|---|---|---|---|
| **Comparison** | Tapir (0.0) | Tapir (0.5) | Tapir (0.9) |
| **Traditional B – *CostMAP* B** | 0.731 (z = 0.74) | 0.021 (z = -1.64) | 0.001 (z = -8.88) |
| **Traditional B – Null Surface** | 0.933 (z = -0.22) | 0.001 (z = 4.73) | 0.001 (z = -4.93) |
| ***CostMAP* B – Null Surface** | 0.808 (z = 5.22) | 0.001 (z = 6.37) | 0.001 (z = 13.81) |

The p-values from the Wilcoxon-Mann-Whitney test for the river corridor scenario are given in Table 2. With a p-value of 0.001 for tapir (0.5), tapir (0.9), and the three raster models (Null Surface, Traditional-Surface A, and *CostMAP*-Surface B), the Kruskal-Wallis test showed a significant difference for the river corridor scenario. With a p-value of 0.749, the Kruskal-Wallis test did not show a significant difference at for tapirs at 0.0 persistence and the three raster models.

The boxplots in Figure 6 visualize the five-number summary of the three raster surface models and how each tapir group responded to the weighting system. As the persistence level increased from 0.0 to 0.9, the difference between how the tapirs responded to the three cost surface models became more pronounced. Tapirs moving along *CostMAP*-Surface designed to aid movement along rivers did use river corridors more often in all scenarios, but the results were only significant at the 0.5 and 0.9 persistence level. In the case of the tapir with a 0.5 and 0.9 persistence level, tapirs moving over the surface created with a queen's kernel also used the corridors more than on the Null Surface. However, the amount of corridor usage on that surface was not significantly different than on the Null Surface for tapir (0.0). Except in the case of tapir (0.0), the high z-scores suggests there is a strong difference between the three surfaces.

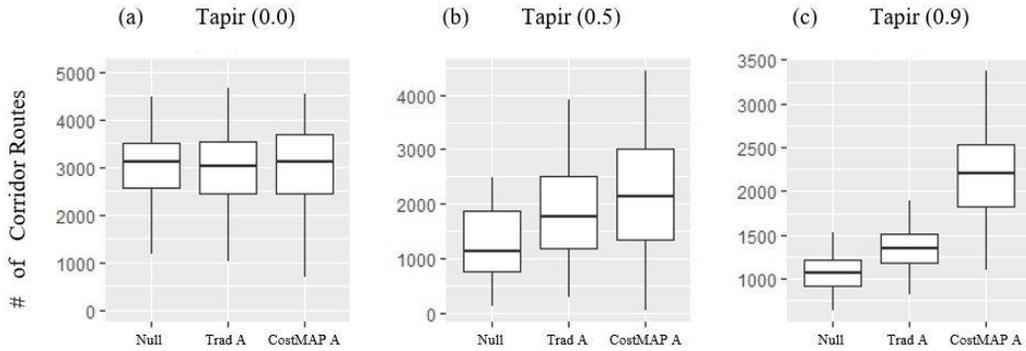

*Figure 6. Boxplots of the number of times the tapir used river corridors (defined as three consecutive steps) when walking over each raster model (Null Surface, Traditional-Surface B, and CostMAP-Surface B).*

# 4 $CO_2$ capture and storage case study

$CO_2$ capture and storage (CCS) can be broken down into a three-step procedure: (1) capture and compression of $CO_2$ before its release to the atmosphere from stationary sources such as power plants; (2) transportation of the captured $CO_2$ through a pipeline network; and (3) injection of the $CO_2$ into deep geological reservoirs such as saline aquifers and depleted oil and gas fields (Middleton and Bielicki 2009). A fourth and optional step, but one that would improve the economic viability of CCS infrastructure, involves using the captured $CO_2$ for other manufacturing processes that produce market-viable products (Middleton et al. 2015), such as in the use of carbon-neutral cementation (Vance et al., 2015).

## 4.1 *SimCCS$^{2.0}$*

The transportation component of CCS requires development of extensive pipeline networks, transporting massive volumes of $CO_2$ over long distances between multiple sources and sinks (Middleton et al. 2014). The pipeline network not only has a significant contribution to costs, the network actually helps drive decisions about where, how much, and when to capture $CO_2$ (Middleton et al. 2012a). Even before the pipeline network is identified, decision makers need to define a large set of potential corridors where pipelines could be constructed, and these corridors critically rely on the underlying cost surface. This pipeline routing and networking problem is addressed in the software *SimCCS$^{2.0}$* which develops a realistic, pipeline network that connects $CO_2$ sources and reservoirs (Middleton and Bielicki 2009). *SimCCS$^{2.0}$* accounts for topographic, social, and geometric costs of a CCS pipeline network to accurately portray the economic feasibility of its deployment within a geographic region. The original *SimCCS$^{2.0}$* software used a cost surface and candidate network approach comprehensively described in Middleton et al. ((2012b) *SimCCS$^{2.0}$* was completely redeveloped in 2018, including an updated mixed integer-linear programming formulation, open-source Java-based coding, and coupled with high-performance computing (Middleton et al. 2018). The latest version of *SimCCS$^{2.0}$* is fully integrated with *CostMAP*.

## 4.2 Data and methods for CCS case study

*SimCCS*[2.0] uses Delaunay triangles to create an initial, idealized network between sources and sinks that is not beholden to real-world environmental or social costs (Figure 7a). *SimCCS*[2.0] then solves a least cost path between each node that is connected via a Delaunay triangle edge using the cost values retrieved from *CostMAP* within Dijkstra's shortest path algorithm (Figure 7b). Additional node pairs can be connected by the user. In the default cost surface, each pixel represents roughly a one square kilometer landscape and has a value that estimates the cost of building a pipeline within that area. A subset of the least costs paths are selected as a candidate pipeline network, which in turn corresponds to the set of possible pipeline routes accouting for factors such as $CO_2$ prices and transport costs.

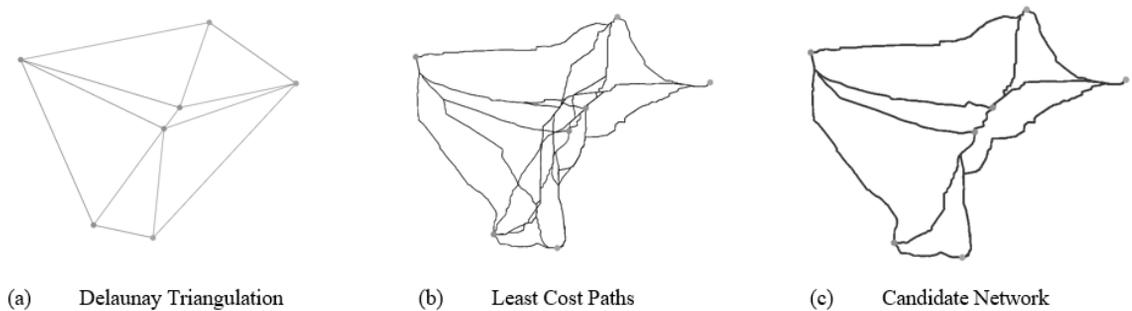

(a) Delaunay Triangulation    (b) Least Cost Paths    (c) Candidate Network

*Figure 7. Illustration of SimCCS*[2.0] *LCP analysis. SimCCS*[2.0] *computes an idealized (c) Candidate Network for CCS pipelines through a combination of (a) Delaunay triangles to find the optimal spatial arrangement and by utilizing a cost surface to compute the (b) least cost path across the cost surface via Dijkstra's algorithm.*

For the CCS case study, we constructed cost surfaces using real-world data that includes costs for land cover, land ownership, topography, population, roads, rivers, railroads, and pipeline data. We used the USGS National Land Cover Database to derive land cover weights (Homer et al. 2015). For road, river, and railway weighting we used the 2011 U.S. Census Tiger Line data (Branch 2011). Topography weights were computed from a USGS DEM that was upscaled to one kilometer (Gesch et al. 2002). Pipeline data was provided by the Pipeline and Hazardous Materials Safety Administration. We created hypothetical carbon sources and sinks in ArcMap 10.5 near rivers, roads, and railroads, so the outcomes in *SimCCS*[2.0] would be forced to account for crossings and corridors. For the cost surfaces, we created a Traditional Cost Surface using a queen's kernel and we created a *CostMAP* Cost Surface using the new multi-scale method available in *CostMAP*. For both the surfaces, we heavily weighted river, road, and railroad crossings as barriers. Next we created surfaces but favored roads and pipelines as corridors. We used both cost surfaces with the same source and sink data in *SimCCS*[2.0] to compute optimal CCS pipelines and compared the results.

## 4.3 Data and methods for CCS case study

Figure 8 demonstrates the optimal path results from surfaces using a queen's kernel or the search kernel in *SimCCS*[2.0]. Pipeline least costs paths tended to cross rivers, roads, and railroads more often over the traditional surface more than the *CostMAP* Surface, despite weighing crossings the same in each cost surface calculation. The *CostMAP* Surface did not completely prevent river crossings—which it should not—but the solutions usually found a shorter path across a river (Figure 8). It's worth noting that *SimCCS*[2.0] uses Delaunay triangulation to find feasible source and sink connections and will find a path between connecting points, even if the cost surface is heavily weighted to prevent crossings (because all sinks and sources need to be connected by at least one feasible corridor). Ideally, however, the LCP analysis will find routes that minimizes the costs of crossing. For example, in Figure 8(c) the calculation of the pipeline network over the traditional surface was more prone to utilize pixels with rivers. To quantify the effectiveness of CostMAP, we used novel approaches that have not before been applied to CCS infrastructure routing. First, across our entire study area, we found that the pipeline network calculated over the *CostMAP* surface utilized pixels that contained barriers (i.e. rivers, roads, and rails) 55% less than on the traditional surface.

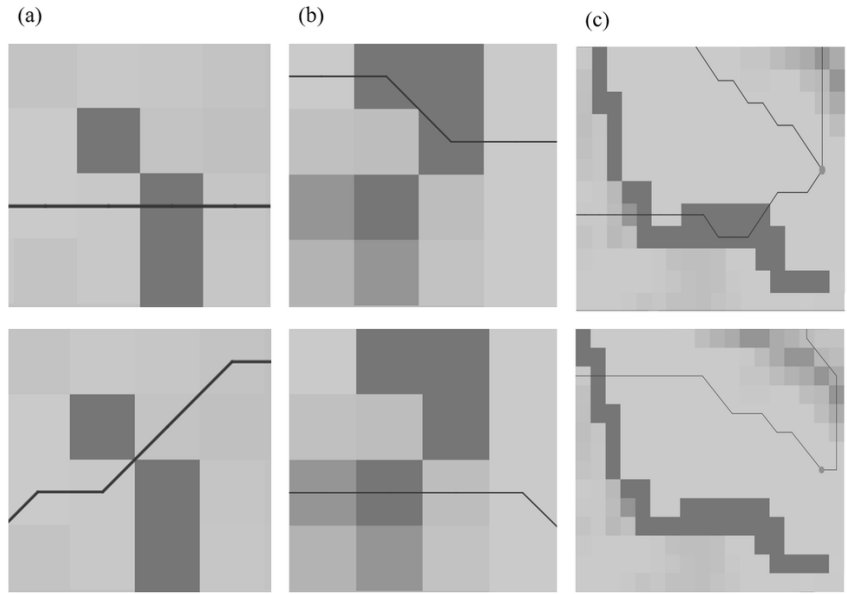

*Figure 8. Example areas comparing river crossings (dark shaded pixels) in SimCCS$^{2.0}$ using the same source and sink data, but with the two different cost rasters. The lighter colors represent areas with no rivers. **Top** shows the results of a calculated pipeline (dark polyline) using the traditional queen's kernel. **Bottom** shows the results from a calculated pipeline (dark polyline) using the search kernel in CostMAP.*

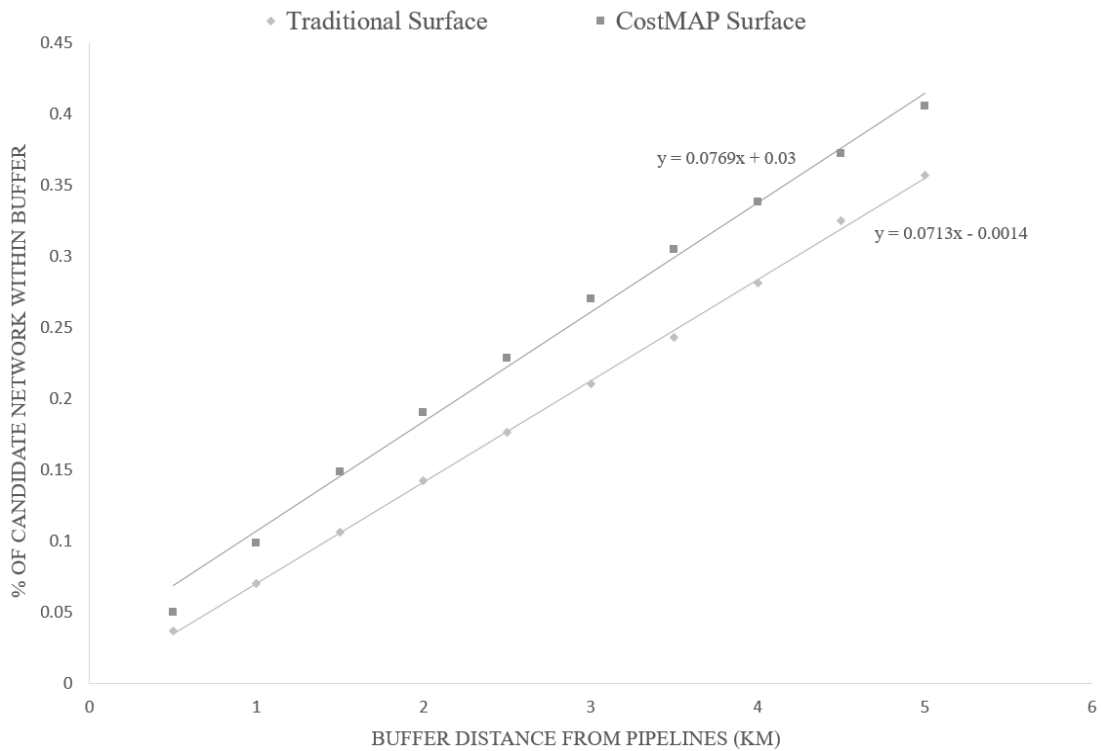

*Figure 9: Regression plot showing the difference between the percentages of the candidate network that fell within a buffer range of 500–5000 meters for networks generated over a CostMAP surface using the search kernel compared to a traditional surface.*

Secondly, to quantify how effective a *CostMAP* surface compared to the traditional methods in weighing linear features as rights-of-way, we calculated how much of the least cost path network generated in *SimCCS*$^{2.0}$ was within 0.5, 1.0, 1.5, 2.0, 2.5, 3.0, 3.5, 4.0, 4.5, and 5.0 kilometers of an already existing pipeline network (Figure 9). With a p-value of 0.001, we found there was as a significant difference at a < 0.05 probability level between how much of the CCS pipeline network ran nearer to an already existing pipeline network (as defined by ROWs) between the two surface types. At all buffer distances the pipeline network generated using the *CostMAP* surface had a greater percentage within the buffer zone. Within 1km of a pipeline network, the network generated using the *CostMAP* surface increased from 6.9% to 9.9%. At a distance of 2 km the percentage of pipeline within the buffer zone increased from 14.1% to 19.0%.

# 5 Discussion

The results from the tapir case study illustrate that cost surfaces created with the new search kernel approach in *CostMAP* perform significantly better at preventing crossings than cost surfaces created with a queen's kernel. The results were less conclusive for the corridor scenario. However, *CostMAP*-Surface B did have a significant greater usage of river corridors for the tapir with persistent levels of 0.5 and 0.9 than did Traditional-Surface B. The use of these persistence levels mimics short- versus long-term goals in movements, where a high persistence level would correspond to reaching a long-term goal. Potentially, longer-term goals are more motivated by corridor use (LaPoint et al. 2013). In that case, the results of this case study are promising, since *CostMAP*-surfaces engender more corridor use for longer-term goals.

Future research for *CostMAP* will include modeling animals of different size and mobilities, and more particular movement behaviors (e.g. foraging, resting, migration, predation, and so on). Future research will also include investigating how *CostMAP* performs for other types and scales of barriers and corridors (e.g., different river sizes, roads of varying traffic conditions, or conservation easements). In addition, other variables, like an animal's energy expenditure can be integrated into *CostMAP* to incorporate how an animal's "energy landscape" impacts its movements (Shepard et al. 2013). Overall, the results from this case study demonstrate how the added accuracy of costs surfaces created in *CostMAP* will be helpful for understanding the nexus between animal movements and the environment.

For the CCS case study, significantly improved routes are identified with the least cost path analysis using corridors more often and finding areas to cross barriers at location that are less costly using the *CostMAP* surface. For instance, Figure 8 illustrates the surfaces created with the queen's kernel were significantly not as accurate in avoiding crossings as those surfaces created with the search kernel.

Future research will include increased testing and quantification of how the cost network created with the search kernel *CostMAP* compares to a network created with a queen's kernel. That increased testing will be possible in the future with the completion of the *SimCCS*$^{2.0}$ *Gateway*, which is a high-performance computing platform that will allow for thousands of *SimCCS*$^{2.0}$ runs. Future research will also include utilizing a knight's kernel in *CostMAP*. However, one of the goals of *SimCCS*$^{2.0}$ is to solve LCP scenarios for CCS at regional or even continental scale. Figure 10 illustrates the *CostMAP* ability to calculate complex national-scale cost surfaces in a matter of minutes. The disadvantage to adding a knight's kernel is that it is more computationally demanding (Bevan 2011). Currently, *CostMAP* creates weighted-cost surfaces and weighted-cost networks in a relatively short time (computational runtime is ~5 minutes for a raster with 6.9 million pixels and ~48 million graph edges). At finer scales, a knight's kernel would be less intensive and could increase accuracy of the cost network. There has been some question about how well knight's kernels improve accuracy (Bevan 2011), however, there are instances (e.g., steep slopes) where a knights kernel has proven more accurate in finding routes that are less costly (Yu, Lee and Munro-Stasiuk 2003). In addition, since our weighting system is not a simple accumutalive cost surface, rather a combination of accumulation, Boolean, and nonlinear decisions (Kuby et al., 2011; Middleton et al., 2012b) a Multi Criteria Decision Analysis (MCDA) approach is difficult. The impact that weighting has on results can be significant, so we have long-term future plans to quantitatively develop weights from actual pipeline projects and relating these to geographic variables in a nonlinear fashion. Currently, our weights were developed from the literature and expert opinion, and not a formal MCDA approach. However, in an ongoing project focused on routing large $CO_2$ pipelines, we will continue to use the current approach particularly with the sensitivity of native lands and other environmental justice variables for pipeline construction. In the near future, we plan to use *CostMAP* as a tool to engage with communities in participant mapping.

Beyond the case studies presented here, *CostMAP* could be used in any number of other projects focused on routing. In particular, the integration of *CostMAP* and *SimCCS*$^{2.0}$ would allow for seamless analysis into other avenues of routing planning (e.g. underground transmission lines, gas and water pipelines), and could be a valuable tool for planning large scale linear infrastructure extensions and maintenance to decrease costs and improve efficiency (Kielhauser 2018).

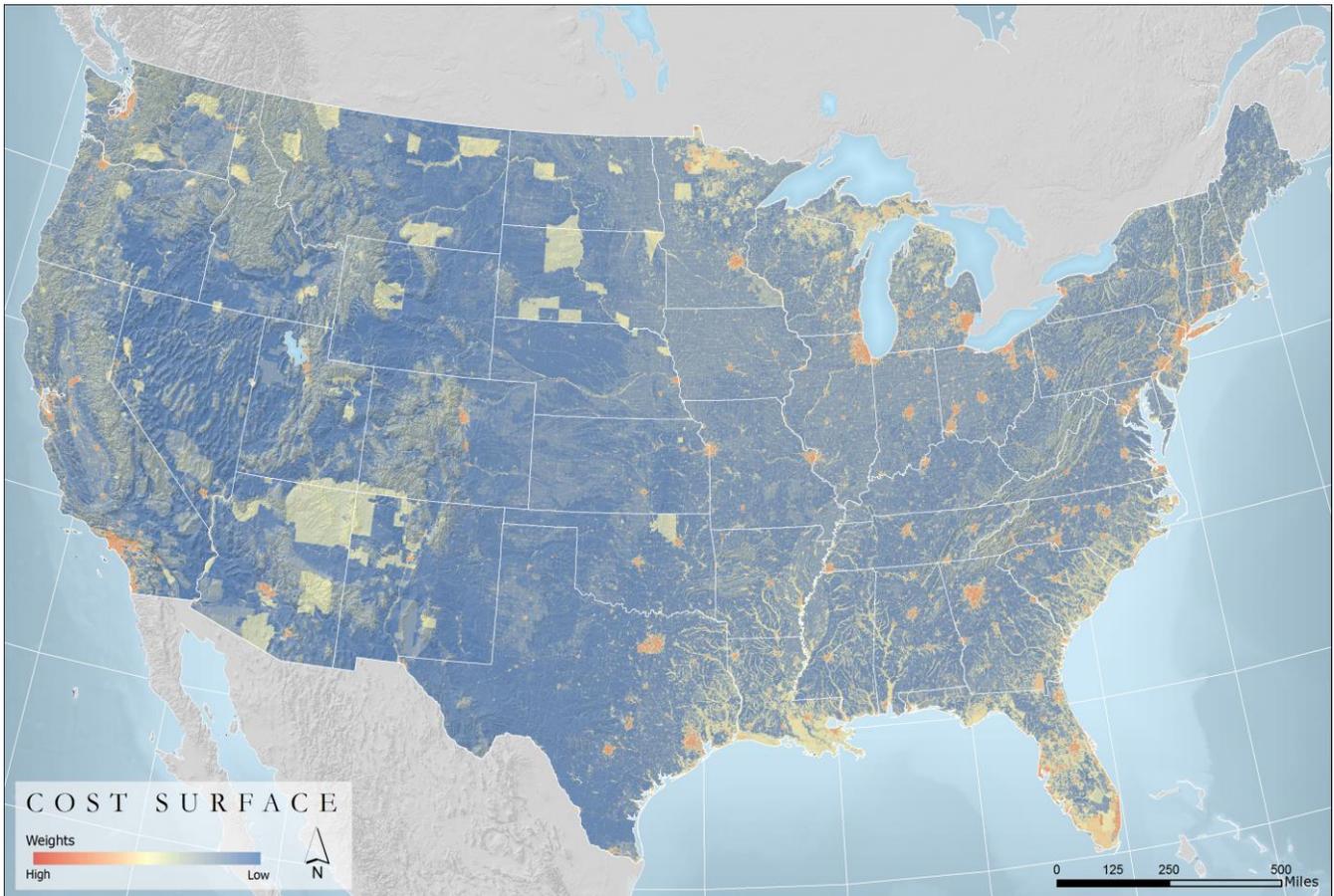

Figure 10: National cost surface produced by CostMAP for the pipeline case. High-cost areas (red) are mostly dense urban areas. Medium-cost areas (yellow) include Native American lands (e.g., Navajo Nation), rivers and wetlands (e.g., Mississippi delta), areas of steep topography (e.g., Rocky Mountains and the Sierra Nevada), and other protected areas. Note that single-cell barriers and corridors are rarely visible in the GIS output due to "mean smearing" (where the cell weight is averaged over eight directions) or where a cell is both a barrier and corridor (cells can act as both simultaneously, which cannot be shown in a 2D raster image). Barriers that are more than one cell wide—such as major river—are clearly visible since mean-smearing is reduced or absent.

# 6 Conclusion

This study addressed a key research gap in cost raster and network generation: suitably representing corridors and barriers and quantitatively measuring their impact. Corridors and barriers are known to play a critical role in routing through cost rasters and our novel approach should significantly improve accuracy. We also introduced an innovative approach to statistically quantify the impact of barriers and corridors in cost surfaces. A key finding from that statistical approach found that pipeline routes generated over a *CostMAP* surface used pixels that contained barriers 55% less than routes created over a traditional GIS cost surface. In fact, in both case studies, the search kernel used in *CostMAP* demonstrated higher accuracy in limiting crossings and utilizing corridors than the traditional queen's kernel in both the raster surface and cost network. Consequently, results from the study demonstrate that our new approach significantly improves performance over traditional ways of presenting corridors and barriers in cost rasters and networks. While our results are promising, we know that how features are weighed can affect

LCP analysis. With our future plans to quantitatively develop weights from existing routing, we aim to remove ambiguity that can arise when weights are assigned based upon literature or expert knowledge.

As part of this research, we also introduced a new tool—*CostMAP*—that automatically incorporates the new corridor-and-barrier approach into costs surfaces and networks. Although the tool itself is not transformative, it does allow researchers and other users to readily represent accurate corridors and barriers for routing. In addition, since *CostMAP* is open source, it can be readily modified to address project specific concerns including custom weighing approaches and new data layers. For example, we are concerned with environmental justice and plan to use *CostMAP* to identify CCS pipeline routes around protected and sensitive areas. We are continuing to develop *CostMAP* and plan to include other features, making it even easier to use for researchers performing LCP analysis. *CostMAP* is available on github at: https://github.com/simccs/costmap.